# Experimental minimum of condensed-phase optical refrigeration

Zhuoming Zhang[1], Yang Ding[1], Peter J. Pauzauskie[2],
Mansoor Sheik-Bahae[3], Denis V. Seletskiy[4], Masaru Kuno[1,5]

1. University of Notre Dame, Department of Chemistry and Biochemistry, Notre Dame, IN 46556, USA
2. Department of Materials Science, University of Washington, Seattle, Washington 98195, USA
3. Department of Physics and Astronomy, University of New Mexico, Albuquerque, New Mexico 87131, USA
4. Department of Engineering Physics, Polytechnique Montréal, Montréal, QC, H3T 1J4, Canada
5. University of Notre Dame, Department of Physics and Astronomy, Notre Dame, IN 46556, USA

**Abstract**

Since the first demonstration of optical refrigeration in a rare-earth-doped glass nearly 30 years ago, the nascent field of laser cooling solids has progressed significantly. It is now possible to demonstrate payload cooling to ~91 K using laser-excited ytterbium-doped fluoride crystals. Realizing lower temperatures, however, requires achieving ultrahigh purities in existing rare earth-doped crystals or establishing new cooling media. For the latter, semiconductors are an obvious choice given higher cooling power densities and predicted cooling floors as low as 10 K. This has stimulated a race to demonstrate the optical refrigeration of a semiconductor. It is therefore timely to systematize the necessary and sufficient experimental minimum criteria for reporting optical refrigeration results to elevate the reliability and reproducibility of current and future optical refrigeration claims.

We distill an optical refrigeration Experimental Minimum (EM) that we propose will standardize the reporting of new cooling results. EM tenets fall into the following categories:

1) *Demonstrations of explicit heating vs cooling*: Reports should show reliable and self-consistent transitions between heating and cooling regimes by tuning laser excitation frequencies around mean emission frequencies ($\bar{\nu}_{em}$).
2) *Optical cooling metrics*: Two critical quantities, external quantum efficiency ($\eta_{EQE}$) and absorption efficiency ($\eta_{abs}$), should be measured and reported.
3) *Thermodynamic consistency*: Cooling time constants and achieved temperature changes must be consistent with thermodynamic constraints imposed by the cooling environment and sample parameters.
4) *Reliable temperature measurements*: Details of the temperature measurement technique, its calibration procedure, and temperature- as well as time resolution should be reported.

All optical refrigeration claims should demonstrate the above four criteria to ensure their reliability and verifiability. We further propose that the EM serve as a guide for reviewing literature claims in the field.

**Introduction**

Condensed phase laser cooling is premised on removing thermal energy from a material through its anti-Stokes photoluminescence (ASPL).[1,2] Since the first 1995 report of ASPL-induced cooling in a $Yb^{3+}$ doped heavy-metal-fluoride glass, numerous experiments now unambiguously confirm condensed phase optical refrigeration as well as cryogenic cooling in rare-earth (RE)-doped glasses and crystals.[3,4] Of note is a record cooling temperature of 91 K achieved in a $Yb^{3+}$-doped yttrium lithium fluoride crystal.[5] Ongoing efforts in material purification and high-purity crystal growth suggest that 50 K temperatures are within reach for RE-doped systems.[6,7]

Unfortunately, the thermal depopulation of higher energy RE atomic levels at low temperatures freezes out their optical cooling cycles. Deep cryogenic applications therefore require changing optical cooling media. To this end, numerous attempts have been made with alternate, condensed phase systems[8,9] with recent focus shifting to semiconductors[10,11] where Fermi-Dirac statistics allow laser cooling down to LO-phonon freezeout temperatures around 10 K.[12]

Optical refrigeration has been attempted in bulk, direct-gap semiconductors such as ZnTe[13] and lead-based, hybrid/all-inorganic perovskites such as methylammonium lead iodide ($CH_3NH_3PbI_3$).[14] It has also been attempted with low-dimensional semiconductors such as GaAs/InGaP heterostructures[10,11,15], GaAs quantum wells[16], CdS nanobelts[17,18,19], single-layer, two-dimensional Ruddlesden-Popper phenylethylene lead iodide [$(C_6H_5C_2H_4NH_3)_2PbI_4$] microcrystals[14], monolayer $WS_2$[20], and cesium lead bromide ($CsPbBr_3$)[21] or core/shell CdSe/CdS[22] nanocrystals. To various degrees, all attempts have been challenged by a combination of factors, originating from the need to produce high-purity, cooling-grade materials.

Crucially, all current semiconductor optical cooling claims are hampered by incomplete experimental details. Often key performance metrics are implied (or assumed), not measured. On rare occasion, results do not conform to known physics.[23,24] Hence, a need exists to standardize the reporting of optical refrigeration claims to ensure verifiable cooling outcomes.

In what follows, we revisit the tenets of optical refrigeration and distill them into a condensed phase optical refrigeration Experimental Minimum (EM). The EM aims to promote rapid progress in the field by standardizing the reporting of cooling results. We strongly believe that demanding research, especially those which rely on the absolute performance characteristics of advanced materials - be it optical refrigeration, room-temperature superconductivity, light transistors, Majorana fermions, etc… require such standardization.[25] **Figure 1** summarizes optical refrigeration's EM tenets, which are discussed in what follows.

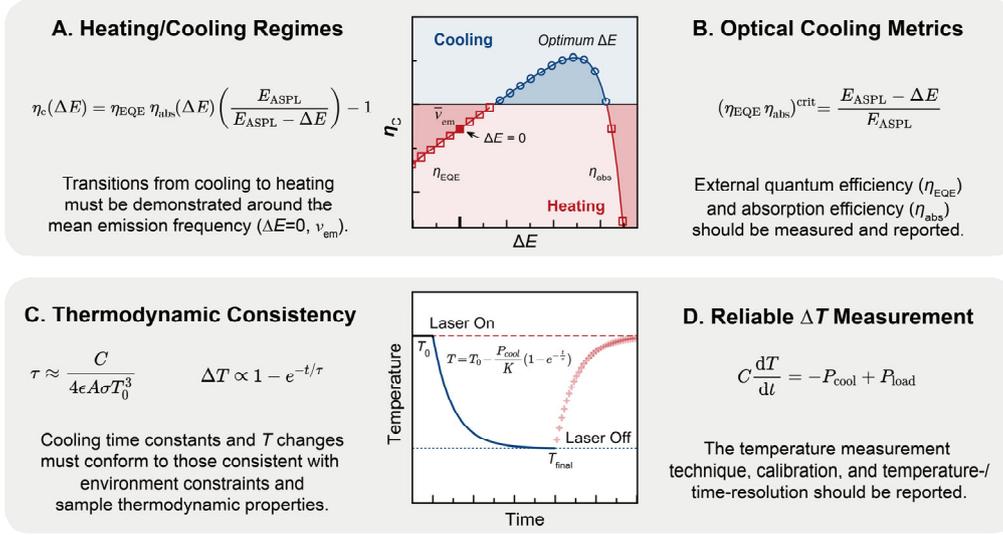

**Figure 1**. Schematic summary of a condensed phase optical refrigeration Experimental Minimum.

## Modeling optical refrigeration

In optical refrigeration, net cooling power density, $P_{cool}$ ($P_{cool} > 0$ represents cooling by the chosen sign convention), is expressed as[1]

$$P_{cool} = P_{em} - P_{abs} \qquad (1)$$

where $P_{abs} = (\alpha + \alpha_b)I_0$ is the absorbed power density for an incident, excitation laser irradiance of $I_0$. $\alpha$ is a resonant, semiconductor absorption coefficient at frequency $\nu_0$ and $\alpha_b$ is a "parasitic" absorption coefficient that accounts for unwanted light absorption, which ultimately leads to heat generation. In RE-doped solids the origin of $\alpha_b$ is believed to be due to trace amounts of transition metal ions, such as $Co^{2+}/Co^{3+}$ and $Fe^{2+}/Fe^{3+}$,[26,27] some of which has been confirmed at the ppm level.[27] For GaAs heterostructures, it has been suggested that $\alpha_b$ stems from defects at GaAs quantum well/InGaP heterojunctions.[7]

In **Equation 1**, the emitted photoluminescence (PL) power density is

$$P_{em} = \alpha I_0 \eta_{EQE} \left( \frac{E_{ASPL}}{E_{ASPL} - \Delta E} \right) \qquad (2)$$

where $\eta_{EQE}$ is the material's external quantum efficiency [also referred to as quantum yield (QY) in the semiconductor nanocrystal community] and $E_{ASPL}$ is the average energy at the mean emission frequency ($\bar{\nu}_{em}$) of the PL. The subscript ASPL reflects the anti-Stokes PL obtained in the laser cooling regime. $E_{ASPL}$ approximately corresponds to the band gap absorption energy, $E_g$, of a direct-gap semiconductor. It can, however, be redshifted relative to it due to excitonic effects or the existence of absorption/emission Stokes shifts[28,29,30], associated with dark exciton[31,32] or optically passive emitting states. To illustrate, in $CsPbBr_3$ nanocrystals with edge lengths between 13 and 4 nm, absorption/emission Stokes shifts range from 20-82 meV, respectively.[29,30] The difference $E_{ASPL} - \Delta E$ thus reflects the detuning ($\Delta E$) of the pump laser into the optical gap relative to the material's mean emission (not absorption) energy.

By combining **Equations 1** and **2**, a cooling efficiency,

$$\boldsymbol{\eta_c} = \frac{P_{\text{cool}}}{P_{\text{abs}}} = \eta_{\text{EQE}} \eta_{\text{abs}} \left(\frac{E_{\text{ASPL}}}{E_{\text{ASPL}} - \Delta E}\right) - 1, \tag{3a}$$

arises where

$$\eta_{\text{EQE}} = \eta_e W_r / (\eta_e W_r + W_{nr}) \tag{3b}$$

and

$$\eta_{\text{abs}} = \alpha/(\alpha + \alpha_b). \tag{3c}$$

In **Equation 3b**, $\eta_{\text{EQE}}$ is defined in terms of band edge, interband radiative ($W_r$) and non-radiative ($W_{nr}$) recombination rates with $\eta_e$ a photon escape efficiency that accounts for light trapping in semiconductors due to refractive index differences with the surrounding medium.[12] For nanocrystals that behave as dipole emitters, $\eta_e \approx 1$.

$\eta_{\text{abs}}$ (**Equation 3c**) is an absorption efficiency, which represents the fraction of the total absorbed pump photons that induce cooling; $1-\eta_{\text{abs}}$ is the fraction resulting in heating via impurity absorption. $\eta_{\text{abs}}$ can also be defined as an up-conversion efficiency that describes the likelihood that a below-gap absorption event leads to up-conversion and possibly to cooling as opposed to non-radiative relaxation.[33,34,35]

Setting $P_{\text{cool}} = 0$ (alternatively, $\boldsymbol{\eta_c} = 0$), reveals a critical dimensionless product $(\eta_{\text{EQE}} \eta_{\text{abs}})^{\text{crit}}$ above which the thermal energy removed by up-converted PL exceeds that gained by non-radiative recombination and parasitic impurity absorption. An associated performance threshold is

$$(\eta_{\text{EQE}} \eta_{\text{abs}})^{\text{crit}} = (E_{\text{ASPL}} - \Delta E)/E_{\text{ASPL}} \tag{4}$$

and captures the challenging nature of solid-state optical refrigeration. Namely, **Equation 4** highlights the interrelationship between non-radiative recombination, photon escape efficiency, and absorption/up-conversion efficiency -- all parameters which dictate permissible $\Delta E$-values for realizing optical cooling. An additional complication arises from $\eta_{\text{EQE}}$'s carrier-density ($N$) dependence that leads to optimal values of $N^{\text{opt}}$ and correspondingly $\eta_{\text{EQE}}^{\text{opt}}$.[12] $\eta_{\text{abs}}$ may also depend on $N$ through absorption saturation given sufficient pumping.

For $\eta_{\text{abs}} = 1$, associated critical $\eta_{\text{EQE}}$-values ($\eta_{\text{EQE}}^{\text{crit}}$) range from 0.85-0.99.[2] For GaAs/InGaP heterostructures and CsPbBr$_3$ nanocrystals with room-temperature optical gaps of 1.42 eV and 2.41 eV, $\eta_{\text{EQE}}^{\text{crit}}$= 0.93 and 0.96 respectively ($\Delta E = 100$ meV).[2] Even realizing $\eta_{\text{EQE}}^{\text{crit}}$ does not guarantee cooling. To illustrate, despite $\eta_{\text{EQE}} > 0.99$ in GaAs, cooling has not been achieved due to $\alpha_b$-induced heating. This prevents **Equation 4** from being satisfied so that $P_{\text{cool}} < 0$.[10]

Because net laser cooling requires $\boldsymbol{\eta_c} > 0$, $\eta_{\text{EQE}}$ and $\eta_{\text{abs}}$ are important material parameters whose numerical values must be near-unity. That near-unity $\eta_{\text{EQE}}$ and $\eta_{\text{abs}}$ are possible can be seen in RE systems where values of $\boldsymbol{\eta_c}$ fall in the range 0-0.05.[5,36] For semiconductors, near-unity $\eta_{EQE}$ and $\eta_{abs}$ require material optimization via a combination of purification, defect passivation, and proper material selection. Systems with large electron-phonon coupling constants are therefore currently believed to be preferential for cooling, as they can possess intrinsically large up-conversion efficiencies.[37]

> *EM 1: Demonstrations of explicit heating vs cooling.* Reproducible transitions from heating to cooling should be demonstrated as functions of $\Delta E$ about a system's mean emission energy.

**Equations 3a** and **4** reveal that changing $\Delta E$ about $\bar{\nu}_{em}$ transitions a system between its cooling and heating regimes. While this can be used to optimize $\boldsymbol{\eta}_c$, more important is that self-consistency requires that a material's cooling and heating regimes be explicitly demonstrated. This can be done in practice using Laser-Induced Thermal Modulation Spectroscopy (LITMoS).[42]

> *EM 2: Optical cooling metrics.* Critical parameters required to cool should be verified and explicitly reported. Report measurements of $\eta_{EQE}$ and $\eta_{abs}$ together with relevant experimental conditions, such as injected $N$, $\Delta E$, $T$, $\alpha_b$ etc…

Experimental $\eta_{EQE}$ measurements involve relative, absolute, model-based, and calorimetric approaches.[2,38] Relative methods involve comparing a specimen's integrated emission spectrum to that of a reference specimen with a known $\eta_{EQE}$. Absolute approaches use an integrating sphere. Model-based approaches such as power-dependent photoluminescence (PDPL)[33] invoke nonlinear $I_0$ emission intensity ($I_{em}$) dependencies and kinetic models to fit $I_{em}$ from where both a maximum $\eta_{EQE}$ and corresponding $N^{opt}$ can be found.

Calorimetric approaches entail registering the temperature of a sample as a function of the excitation laser frequency, $\nu_0$, such that $\eta_{EQE}$ can be precisely read off from the slope, $dT/d\nu_0$, of measurements in the region where $h\nu_0 > E_{ASPL}$[36], see also **Figure 1**. These temperature measurements commonly employ one or several non-contact techniques, including thermal imaging, calibrated temperature-dependent emission spectra, or photothermal deflection.[2,39]

A further advantage of calorimetric measurements is simultaneous estimates of $\alpha_b$ when $h\nu_0 < E_{ASPL}$. For CsPbBr$_3$ nanocrystals, $\eta_{abs} = 0.75$ ($\Delta E = 23$ meV) has been measured via direct measurements of Stokes and anti-Stokes excitation irradiances required to achieve identical Stokes/ASPL emission intensities in conjunction with independent estimates of corresponding Stokes and anti-Stokes absorptances.[34] More extensive measurements across a wider range of temperatures and $\Delta E$-values now reveal $\eta_{abs} \sim 1$ in CsPbBr$_3$ nanocrystals.[35]

> *EM 3: Thermodynamic consistency.* Cooling results should be explicitly compared to thermodynamically-predicted cooling timescales and final temperatures.

Reported cooling results must furthermore be consistent with thermodynamic considerations of the specimen and its environment[40] wherein

$$C\frac{dT}{dt} = -P_{\text{cool}}(T) + P_{\text{load}}(T)$$
$$\approx -P_{\text{cool}} + \left[4\epsilon A\sigma T_0^3 + \kappa_v A + \frac{\kappa_c A_c}{d_c}\right](T_0 - T)$$
$$\approx -P_{\text{cool}} + K(T_0 - T). \quad (5)$$

**Equation 5** expresses a material's temperature change due to a competition between optical refrigeration and heating from extrinsic thermal loads ($P_{\text{load}}$). The first bracketed term in $P_{\text{load}}$ represents blackbody heating from a specimen's surroundings (*e.g.*, a vacuum chamber). The second and third terms represent environmental convective and conductive loads on the specimen. $C$ is the sample's heat capacity, $\epsilon$ is the specimen chamber's emissivity, $A$ is the semiconductor's surface area, $\sigma$ is the Stephan-Boltzmann coefficient, $T_0$ is an initial temperature, $\kappa_v$ is a convective heat transfer coefficient, $\kappa_c$ is a thermal conductivity constant, $A_c$ is a contact area, and $d_c$ is a contact point length.

**Equation 5** yields a time-dependent temperature ($T$) and corresponding cooling time constant, $\tau$. In the limit convective and conductive thermal loads are negligible and where $\Delta T = T_0 - T \ll T_0$,

$$T(t) \approx T_0 - \left(P_{\text{cool}}/K\right)(1 - e^{-t/\tau}),$$

with
$$\tau = \frac{C}{K} \approx \frac{C}{4\epsilon A\sigma T_0^3}. \quad (6)$$

**Equation 6** can therefore be compared to experimentally-observed cooling timescales to validate their appropriateness. Heating timescale should likewise be consistent with $P_{\text{load}}$.[23] For reference, prior estimates for CsPbBr$_3$ nanocrystals embedded in an aerogel disk yield $\tau \sim 0.5$ seconds.[2] For an individual CdS nanobelt suspended in vacuum, a maximum time constant is $\tau \sim 30$ ms.[23]

> *EM 4: Reliable temperature measurements.* Details of the temperature measurement technique, its calibration procedure, and temperature- as well as time-resolution should be reported.

Critical to assessing a measurement's conformity with **Equation 6** are accurate time-resolved or steady-state measurements of $T$. In practice, steady-state measurements are carried out upon reaching thermal equilibrium with achieved final temperatures, $T_{\text{final}}$, measured using non-contact (optical) approaches such as up-conversion emission thermometry [41], differential luminescence thermometry[4], pump-probe luminescence thermometry[17] or via a calibrated thermal camera.[42] Time-dependent $T$ measurements on sub-millisecond[40] to 10s of picosecond timescales have also been devised, the latter allowing the first observation of transient laser cooling in GaAs at room temperature.[43]

Extracted temperatures should ultimately be compared to $T_{\text{final}} = T_0 - \frac{P_{\text{cool}}}{K}$ from **Equation 6**.

In summary, there exists much promise to advance condensed phase optical refrigeration beyond RE systems. Possible semiconductor candidates include GaAs [43] and novel nanostructures that exhibit near-unity up-conversion efficiencies.[2,34,35] However, the realization of tangible advances is premised on standardizing reported cooling claims to the above-outlined condensed-phase EM.

**Acknowledgements**

This perspective is borne out of the vision that Mansoor Sheik-Bahae shared with his co-authors a few months before passing away from cancer in July 2023. The article is dedicated to his memory. Financial support from the MURI:MARBLe project, funded by the Air Force Office of Scientific Research (Award No. FA9550-16-1-0362), is acknowledged.

**Competing interests**
The authors declare no competing interests.

**References**

[1] Sheik-Bahae, M. & Epstein, R. I. Laser cooling of solids. *Laser Photon. Rev.* **3**, 67-84 (2009).
[2] Zhang, S., Zhukovskyi, M., Jankó, B. & Kuno, M. Progress in laser cooling semiconductor nanocrystals and nanostructures. *NPG Asia Mater.* **11**, 54 (2019)
[3] Epstein, R. I., Buchwald, M. I., Edwards, B. C., Gosnell, T. R. & Mungan, C. E. Observation of laser-induced fluorescent cooling of a solid. *Nature* **377**, 500-503 (1995).
[4] Seletskiy, D. V. *et al.* Laser cooling of solids to cryogenic temperatures. *Nat. Photonics* **4**, 161-164 (2010).
[5] Melgaard, S. D., Albrecht, A. R., Hehlen, M. P. & Sheik-Bahae, M. Solid-state optical refrigeration to sub-100 Kelvin regime. *Sci. Rep.* **6**, 20380 (2016)
[6] Volpi, A. *et al.* Optical refrigeration: the role of parasitic absorption at cryogenic temperatures. *Opt. Express* **27**, 29710 (2019).
[7] Giannini, N., Yang, Z., Albrecht, A. R. & Sheik-Bahae, M. Investigation into the origin of parasitic absorption in GaInP|GaAs double heterostructures. *SPIE Proc. Optical and Electronic Cooling of Solids II* **10121**, 101210F (2017).
[8] Clark, J. L. & Rumbles, G. Laser cooling in the condensed phase by frequency up-conversion. *Phys. Rev. Lett.* **76**, 2037-2040 (1996).
[9] Clark, J. L., Miller, P. F. & Rumbles, G. Red edge photophysics of ethanolic rhodamine 101 and the observation of laser cooling in the condensed phase. *J. Phys. Chem. A* **102**, 4428-4437 (1998).
[10] Bender, D. A., Cederberg, J. G., Wang, C. & Sheik-Bahae, M. Development of high quantum efficiency GaAs/GaInP double heterostructures for laser cooling. *Appl. Phys. Lett.* **102**, 252102 (2013).
[11] Gauck, H., Gfroerer, T. H., Renn, M. J., Cornell, E. A. & Bertness, K. A. External radiative quantum efficiency of 96% from a GaAs / GaInP heterostructure. *Appl. Phys. A Mater. Sci. Process.* **64**, 143-147 (1997).

[12] Sheik-Bahae, M. & Epstein, R. I. Can laser light cool semiconductors? *Phys. Rev. Lett.* **92**, 247403 (2004)

[13] Zhang, J., Zhang, Q., Wang, X., Kwek, L. C. & Xiong, Q. Resolved-sideband Raman cooling of an optical phonon in semiconductor materials. *Nat. Photonics* **10**, 600-605 (2016).

[14] Ha, S.-T., Shen, C., Zhang, J. & Xiong, Q. Laser cooling of organic–inorganic lead halide perovskites. *Nat. Photonics* **10**, 115-121 (2016).

[15] Giannini, N., Yang, Z., Rai, A. K., Albrecht, A. R. & Sheik-Bahae, M. Near-unity external quantum efficiency in GaAs|AlGaAs heterostructures grown by molecular beam epitaxy. *Phys. Status Solidi RRL* **15**, 2100106 (2021)

[16] Finkeißen, E., Potemski, M., Wyder, P., Viña, L. & Weimann, G. Cooling of a semiconductor by luminescence up-conversion. *Appl. Phys. Lett.* **75**, 1258-1260 (1999).

[17] Zhang, J., Li, D., Chen, R. & Xiong, Q. Laser cooling of a semiconductor by 40 kelvin. *Nature* **493**, 504-508 (2013).

[18] Li, D., Zhang, J. & Xiong, Q. Laser cooling of CdS nanobelts: thickness matters. *Opt. Express* **21**, 19302-19310 (2013).

[19] Li, D., Zhang, J., Wang, X., Huang, B. & Xiong, Q. Solid-state semiconductor optical cryocooler based on CdS nanobelts. *Nano Lett.* **14**, 4724-728 (2014).

[20] Lai, J.-M., Sun, Y.-J., Tan, Q.-H., Tan, P.-H. & Zhang, J. Laser Cooling of a Lattice Vibration in van der Waals Semiconductor. *Nano Lett.* **22**, 7129-7135 (2022).

[21] Roman, B. J., Villegas, N. M., Lytle, K. & Sheldon, M. Optically cooling cesium lead tribromide nanocrystals. *Nano Lett.* **20**, 8874-8879 (2020).

[22] Ye, Z. *et al*. Phonon-assisted up-conversion photoluminescence of quantum dots. *Nat. Commun.* **12**, 4283 (2021).

[23] Morozov, Y. V. *et al.* Can lasers really refrigerate CdS nanobelts? *Nature* **570**, E60-E61 (2019).

[24] Mungan, C. E. & Gosnell, T. R. Comment on "Laser Cooling in the Condensed Phase by Frequency Up-Conversion". *Phys. Rev. Lett* **77**, 2840-2840 (1996).

[25] Checklists work to improve science. *Nature* **556**, 273-274 (2018).

[26] Hehlen, M. P., Epstein, R. I. & Inoue, H. Model of laser cooling in the $Yb^{3+}$-doped fluorozirconate glass ZBLAN. *Phys. Rev. B* **75**, 144302 (2007).

[27] Melgaard, S., Seletskiy, D., Polyak, V., Asmerom, Y. & Sheik-Bahae, M. Identification of parasitic losses in Yb:YLF and prospects for optical refrigeration down to 80K. *Opt. Express* **22**, 7756 (2014).

[28] Kuno, M., Lee, J. K., Dabbousi, B. O., Mikulec, F. V. & Bawendi, M. G. The band edge luminescence of surface modified CdSe nanocrystallites: Probing the luminescing state. *J. Chem. Phys.* **106**, 9869-9882 (1997).

[29] Brennan, M. C. *et al.* Origin of the size-dependent Stokes shift in $CsPbBr_3$ perovskite nanocrystals. *J. Am. Chem. Soc.* **139**, 12201-12208 (2017).

[30] Brennan, M. C. *et al.* Universal size-dependent stokes shifts in lead halide perovskite nanocrystals. *J. Phys. Chem. Lett.* **11**, 4937-4944 (2020).

[31] Efros, A. L. *et al.* Band-edge exciton in quantum dots of semiconductors with a degenerate valence band: Dark and bright exciton states. *Phys. Rev. B* **54**, 4843-4856 (1996).

[32] Sercel, P. C. *et al.* Exciton fine structure in perovskite nanocrystals. *Nano Lett.* **19**, 4068-4077 (2019).

[33] Morozov, Y. V. *et al.* Defect-mediated CdS nanobelt photoluminescence up-conversion. *J. Phys. Chem. C* **121**, 16607-16616 (2017).

[34] Morozov, Y. V., Zhang, S., Brennan, M. C., Janko, B. & Kuno, M. Photoluminescence up-conversion in CsPbBr$_3$ nanocrystals. *ACS Energy Lett.* **2**, 2514-2515 (2017).

[35] Zhang, Z. *et al.* Resonant multiple-phonon absorption causes efficient anti-stokes photoluminescence in CsPbBr$_3$ nanocrystals. *ACS Nano* **18**, 6438-6444 (2024).

[36] Seletskiy, D. V. *et al.* Precise determination of minimum achievable temperature for solid-state optical refrigeration. *J. Lumin.* **133**, 5-9 (2013).

[37] Yamada, Y. & Kanemitsu, Y. Electron-phonon interactions in halide perovskites. *NPG Asia Mater.* **14**, 48 (2022)

[38] Brouwer, A. M. Standards for photoluminescence quantum yield measurements in solution (IUPAC Technical Report). *Pure Appl. Chem.* **83**, 2213-2228 (2011).

[39] Seletskiy, D. V., Epstein, R. & Sheik-Bahae, M. Laser cooling in solids: advances and prospects. *Rep. Prog. Phys.* **79**, 096401 (2016).

[40] Seletskiy, D. V., Melgaard, S. D., Di Lieto, A., Tonelli, M. & Sheik-Bahae, M. Laser cooling of a semiconductor load to 165 K. *Opt. Express* **18**, 18061-18066 (2010).

[41] Zhang, S., Zhang, Z., Zhukovskyi, M., Jankó, B. & Kuno, M. Up-conversion emission thermometry for semiconductor laser cooling. *J. Lumin.* **222**, 117088 (2020).

[42] Mobini, E. *et al.* Laser cooling of ytterbium-doped silica glass. *Commun. Phys.* **3**, 134 (2020)

[43] Lippmann, J. F., Leitenstorfer, A. & Seletskiy, D. V. Laser cooling of semiconductors traced in the time domain. in *Conference on Lasers and Electro-Optics* FW4M.3. (OSA, Washington, D.C., 2019).